\documentclass[twocolumn,iop]{emulateapj}

\usepackage{natbib}
\usepackage{apjfonts}
\usepackage{subfigure}
\usepackage{url}
\usepackage[utf8]{inputenc}
\usepackage{fancyref}
\usepackage{hyperref}
\hypersetup{colorlinks=true,allcolors=blue,pdfborder={0 0 0}}

\shorttitle{Comprehensive analysis of HD~105}
\shortauthors{J.~P. Marshall et al.}

\begin{document}

\title{Comprehensive analysis of HD~105, a young Solar System analog}

\author{J.~P. Marshall\altaffilmark{1}}
\affil{Academia Sinica Institute of Astronomy and Astrophysics, 11F of AS/NTU Astronomy-Mathematics Building,\\No.1, Sect. 4, Roosevelt Rd, Taipei 10617, Taiwan, R.O.C.}

\author{J. Milli}
\affil{European Southern Observatory (ESO), Alonso de C\'ordova 3107, Vitacura, Casilla 19001, Santiago, Chile}

\author{\'E. Choquet\altaffilmark{2,3}}
\affil{Department  of  Astronomy,  California  Institute  of  Technology, 1200 E. California Blvd, MC 249-17, Pasadena, CA 91125}

\author{C. del Burgo} 
\affil{Instituto Nacional de Astrof\'isica, \'Optica y Electr\'onica, Luis Enrique Erro 1, Sta. Ma. Tonantzintla, Puebla, Mexico}

\author{G.~M. Kennedy}
\affil{Department of Physics, University of Warwick, Coventry CV4 7AL, UK}

\author{L. Matr\`a}
\affil{Harvard-Smithsonian Center for Astrophysics, 60 Garden Street, Cambridge, MA 02138, USA}

\author{S. Ertel}
\affil{Steward Observatory, University of Arizona, 933 N Cherry Ave, Tucson, AZ 85719, USA}

\author{A. Boccaletti}
\affil{LESIA, Observatoire de Paris, PSL Research University, CNRS, Sorbonne Universit\'es, UPMC Univ. Paris 06, Univ. Paris Diderot, Sorbonne Paris Cit\'e}

\altaffiltext{1}{Computational Engineering and Science Research Centre, University of Southern Queensland, Toowoomba, QLD 4350, Australia}
\altaffiltext{2}{Jet Propulsion Laboratory, California Institute of Technology, 4800 Oak Grove Drive, Pasadena, CA 91109, USA}
\altaffiltext{3}{\textit{Hubble} Fellow}

\begin{abstract}
HD~105 is a nearby, pre-main sequence G0 star hosting a moderately bright debris disc ($L_{\rm dust}/L_{\star} \sim 2.6\times10^{-4}$). HD~105 and its surroundings might therefore be considered an analogue of the young Solar System. We refine the stellar parameters based on an improved Gaia parallax distance, identify it as a pre-main sequence star {with an age of 50~$\pm$~16~Myr}. The circumstellar disc was marginally resolved by \textit{Herschel}/PACS imaging at far-infrared wavelengths. Here we present an archival ALMA observation at 1.3~mm, revealing the extent and orientation of the disc. We also present \textit{HST}/NICMOS and VLT/SPHERE near-infrared images, where we recover the disc in scattered light at the $\geq$~5-$\sigma$ level. This was achieved by employing a novel annular averaging technique, and is the first time this has been achieved for a disc in scattered light. Simultaneous modelling of the available photometry, disc architecture, and detection in scattered light allow better determination of the disc's architecture, and dust grain minimum size, composition, and albedo. We measure the dust albedo to lie between 0.19 and 0.06, the lower value being consistent with Edgeworth-Kuiper belt objects. 
\end{abstract}


\section{Introduction}
\label{sec:intro}

Through the results of the \textit{Kepler} transit-based exoplanet survey, we are now aware that exoplanets are near-ubiquitous \citep{2016Coughlin,2016ForemanMackey}. Circumstellar debris, most commonly identified through excess emission at infrared wavelengths \citep{2014Matthews}, is less frequently found. However, this is more severely biased and limited by current instrumental sensitivity. A detection rate of 20 \% has been recorded for cool discs around nearby FGK-type stars \citep{2013Eiroa,2016Montesinos}, with a slightly higher detection rate around A-type stars \citep{2014Thureau}. A tentative trend for a higher incidence of debris discs is seen around stars with sub-Solar metallicities hosting low mass planets \citep{2012Wyatt,2014aMarshall}. For high mass planets, a recent study of disc-host systems has identified a tentative correlation between the presence of $> 5~M_{\rm Jup}$ planets and debris discs \citep{2017Meshkat}. A study of exoplanet host stars and debris discs reveals  a trend between these components of planetary systems being seen together \citep{2014Matthews}. However, no evidence of correlations between these properties has been seen in larger stellar samples, whilst potentially attributable to the paucity of information on both faint debris discs and low mass planets around nearby stars, the presence of real correlations cannot be ruled out due to sample construction \citep{2015MoroMartin}.

A critical component in developing an understanding of the diversity of architectures exhibited by known planetary systems is to obtain multi-wavelength resolved images of their dusty debris discs \citep[e.g.][]{2014Ertel,2014bMarshall,2016Marshall,2017Hengst}. Using a sample of 34 resolved debris discs a relationship between stellar luminosity and dust properties has been identified, showing that the dust grains around higher luminosity stars are closer to emitting like blackbodies \citep{2014Pawellek}. Fitting that observed relationship with a range of dust material compositions further identified that a mixture of astronomical silicate and water ice provided the best fit to the distribution of spatially resolved discs \citep{2015PawKri}. The presence of icy material in cool debris discs has also been inferred through modelling of continuum emission of several spatially resolved debris discs, supporting the adoption of icy materials in the analysis of debris dust \citep{2012Lebreton,2016Morales}. More recently, a relationship between stellar luminosity and disc radius has been measured for a similar sized sample of debris discs imaged at sub-millimetre wavelengths \citep{2018bMatra}. However, no evidence was found for the trend of disc radius relative to blackbody radius identified at far-infrared wavelengths. This may be attributed to the low spatial resolution and bias toward higher luminosity stars of the far-infrared sample making it less representative overall.

Analysis of the infrared continuum emission from debris dust can only take our understanding so far. To obtain insight into the dust grain composition and structure (porosity) we require measurement of the silicate features at mid-infrared wavelengths \citep[e.g.][]{2006Beichman,2006Chen,2009Lawler,2012Johnson,2015Mittal}, and/or {measurement of scattered light from the disc (either total intensity or polarised light). A review of recent advancements in the interpretation of scattered light imaging of debris discs is presented in \cite{2018Hughes}}. In the case of silicate features, a minority of debris disc host stars exhibit features in their mid-infrared spectra \citep{2006Beichman,2014Chen}, suggesting the constituent dust grains are large ($>~10~\mu$m) and/or cold. In the latter case, the scattered light brightness is poorly correlated with expectations from the disc brightness in continuum emission \citep[e.g.][]{2014Schneider} but, for those discs that have been imaged in scattered light, determination of the dust optical and scattering properties has been possible \citep[e.g.][]{2007Graham,2010Krist,2012Rodigas,2014Rodigas,2014Soummer,2016Choquet,2016Schneider}. Complementary to the properties of solid material in debris discs, the recent detections of a gaseous component to some of these systems \citep[][]{2016Greaves} provides insight into the volatile content of the dust parent bodies \citep[e.g.][]{2017Matra}, and its origins \citep{2017Kral}.

For Sun-like stars, from which we may draw parallels with the evolution of the Solar System, there exist only a few cases with comprehensive multi-wavelength data sets to support such detailed analyses, e.g. HD~15115 \citep{2008Debes,2015Macgregor}, HD~61005 \citep{2007Hines,2009Maness,2010Buenzli,2016Olofsson}, HD~107146 \citep{2011Ertel,2015Ricci,2018Marino}, HD~207129 \citep{2010Krist,2011Marshall,2012Loehne}, HIP~17439 \citep{2014Ertel,2014Schuppler}, and HD~10647 \citep{2010Liseau,2016Schuppler}. The addition of more targets to this list, particularly within the narrow range of stellar spectral types representative of Solar analogues (i.e. early G-type main sequence stars cf a broad range of F6 to K0 spectral types), is therefore important to understand the diversity of potential outcomes for planet formation processes. The relative novelty of the Solar System may therefore be determined by comparison to the range of properties exhibited by analogous debris discs systems. 

HD~105 is a young, Sun-like star at a distance of 40~pc that hosts a moderately bright debris disc ($L_{\rm dust}/L_{\star} \sim 2.6\times10^{-4}$). HD~105 is a member of the Tucana-Horologium association and has a well established age of 28~$\pm$~4~Myr \citep{2000Torres,2000ZucWeb}. Its debris disc was reported as being marginally resolved in far-infrared \textit{Herschel}/PACS imaging observations and found to have a radius of $\sim$~50~au \citep{2012Donaldson}. With a spectral type of G0, HD~105 might therefore be considered an analogue of the young Solar System.

Here we model the continuum emission and structure of HD~105's debris disc using a combination of far-infrared and sub-millimetre photometry obtained from a new reduction of archival \textit{Herschel}/PACS and SPIRE images, and a spatially resolved archival image of the disc at millimetre wavelengths from ALMA. We complement this approach with analysis of near-infrared images from \textit{HST}/NICMOS and VLT/SPHERE in order to constrain the dust optical properties through scattered light.

In Sect. \ref{sec:obs}, we present the imaging and photometric data compiled to model both the stellar and disc components of this system. In Sect. \ref{sec:meth_res} we lay out our approach to analysing the combined data set and state our findings. Sect. \ref{sec:dis} the results are put in context, and their impact on our understanding of this system. Finally, in Sect. \ref{sec:con}, we summarise our findings and detail the conclusions of this work.

\section{Observations}
\label{sec:obs}

We have compiled broad band photometric data from optical to millimetre wavelengths to construct the SED of HD~105. These previous observations, combined with the ALMA, VLT/SPHERE, and \textit{HST}/NICMOS imaging data presented here, facilitate comprehensive modelling of the HD~105 system. A summary of the photometry used in the modelling process is presented in Table \ref{table:disc_phot}.

At optical wavelengths we use the Str\"omgren data from \cite{2015Paunzen}, Johnson $BV$ and Cousins $I$ data from \citep{1997Mermilliod}, near-infrared 2MASS $JHK_{s}$ \citep{2003Cutri}, and \textit{WISE} W1 and W2 fluxes \citep{2010Wright} to scale the model stellar photosphere and gauge its contribution to the total emission. 

In the mid-infrared, the \textit{WISE} W3 (12~$\mu$m) and W4 (22~$\mu$m) fluxes are complemented by \textit{AKARI}/IRC 9~$\mu$m data point \citep{2010Ishihara}, a \textit{Spitzer}/MIPS measurement at 24~$\mu$m, and \textit{Spitzer}/IRS photometry at 13~$\mu$m and 31~$\mu$m. The \textit{Spitzer}/IRS spectrum was obtained from CASSIS\footnote{The Cornell Atlas of Spitzer/IRS Sources (CASSIS) is a product of the Infrared Science Center at Cornell University, supported by NASA and JPL.} \citep{2011Lebouteiller}. As the target exhibits no evidence of mid-infrared excess, the spectrum was scaled to the photospheric model using a least-squares fit weighted by the measurement uncertainties at wavelengths $< 15~\mu$m. Target fluxes were then extracted in two windows centred at 13 and 31~$\mu$m and 4~$\mu$m in width using the error-weighted average of values within the bin.

At far-infrared wavelengths, we combine \textit{ISO}/ISOPHOT measurements at 60 and 90~$\mu$m \citep{2001Spangler} and a \textit{Spitzer}/MIPS measurement at 70~$\mu$m \citep{2009Carpenter} with \textit{Herschel}/PACS measurements at 70, 100, and 160~$\mu$m. The spatial resolution of \textit{ISO} is much poorer (1.5\arcmin~beam FWHM) than \textit{Spitzer} or \textit{Herschel} (6--18\arcsec~beam FWHM); inclusion of background contamination and errors in the zero level calibration elevating the \textit{ISO}-measured fluxes are possibilities \citep{2003Heraudeau,2003delBurgo}. However, the target is well isolated in the \textit{Herschel} maps and there is good agreement between measurements by all three facilities at similar wavelengths so we therefore do not consider it to have a significant impact on the shape of the SED.

The \textit{Herschel}\footnote{\textit{Herschel} is an ESA space observatory with science instruments provided by European-led Principal Investigator consortia and with important participation from NASA.} observations were first presented in \cite{2012Donaldson}. Here we re-reduced these data to obtain revised values for the source fluxes based on updates to the instrument calibration made available in the intervening time. The data reduction and analysis were carried out in the same manner as described in \cite{2013Eiroa}. The PACS data were reduced in the \textit{Herschel} Interactive Processing Environment\footnote{{\sc hipe} is a joint development by the Herschel Science Ground Segment Consortium, consisting of ESA, the NASA Herschel Science Center, and the HIFI, PACS and SPIRE consortia.} \citep[{\sc hipe},][]{2010Ott} using the standard pipeline processing scripts ({\sc hipe} version 15, PACS calibration 78). Target fluxes were measured in each mosaic using an aperture of 15\arcsec radius and a sky annulus of 30 to 40\arcsec. Appropriate aperture and colour corrections (determined by a blackbody fit to the aperture corrected measurements), based on values tabulated in \cite{2014Balog}, were applied to the measurements before modelling. 

We also present sub-millimetre measurements taken from \textit{Herschel}/SPIRE (Programme ot2\_aroberge\_3, PI: A. Roberge) and an APEX/LABOCA measurement at 880~$\mu$m \citep{2010Nilsson}. The SPIRE observations were again reduced in {\sc hipe} version 15, using SPIRE calibration 14\_3. The target fluxes in the three SPIRE maps were measured using the \textit{sourceExtractorSussextractor} routine. The millimetre SED is further constrained by ALMA photometry at 1300~$\mu$m (Programme 2012.1.00437.S, PI: D. Rodriguez) and an ATCA measurement at 9~mm taken from \cite{2017Marshall}.

\begin{deluxetable}{lrlc}
%\centering
\tablewidth{0.45\textwidth}
\tablecolumns{4}
\tablecaption{Photometry used in disc modelling. \label{table:disc_phot}}
\tablehead{
\colhead{Wavelength} & \colhead{Flux}  & \colhead{Instrument/} & \colhead{Ref.} \\
\colhead{($\mu$m)}   & \colhead{(mJy)} & \colhead{Filter}      & \colhead{}     \\
}
\startdata
0.349 &  891~$\pm$~2 & Str\"omgren $u$ & 1 \\
0.411 & 2071~$\pm$~6 & Str\"omgren $b$ & 1 \\
0.440 & 1897~$\pm$~14 & Johnson $B$     & 2 \\
0.467 & 2989~$\pm$~6 & Str\"omgren $v$ & 1 \\
0.546 & 3669~$\pm$~2 & Str\"omgren $y$ & 1 \\
0.550 & 3499~$\pm$~26 & Johnson $V$     & 2 \\
0.790 & 4468~$\pm$~41 & Cousins $I$     & 3 \\
1.235 & 4139~$\pm$~73 & 2MASS $J$       & 4 \\
1.662 & 3425~$\pm$~73 & 2MASS $H$       & 4 \\
2.159 & 2383~$\pm$~42 & 2MASS $K_{s}$   & 4 \\
3.40  & 1152~$\pm$~122 & \textit{WISE} W1 & 5 \\
4.60  &  699~$\pm$~24 & \textit{WISE} W2 & 5 \\
9.00  &  189~$\pm$~13 & AKARI/IRC9       & 6 \\
12.0  &  114~$\pm$~6 & \textit{WISE} W3 & 5 \\
22.0  &   35~$\pm$~2 & \textit{WISE} W4 & 5 \\
24.0  &   29~$\pm$~1 & \textit{Spitzer}/MIPS  & 7 \\
31.0  &   22~$\pm$~7 & \textit{Spitzer}/IRS   & 8 \\
60.0  &  143~$\pm$~3 & \textit{ISO}/PHT   & 9 \\
70.0  &  141~$\pm$~10 & \textit{Spitzer}/MIPS  & 7 \\
70.0  &  132~$\pm$~8 & \textit{Herschel}/PACS & 10 \\
90.0  &  167~$\pm$~8 & \textit{ISO}/PHT   & 9 \\
100.0 &  168~$\pm$~8 & \textit{Herschel}/PACS & 10 \\
160.0 &  112~$\pm$~9 & \textit{Herschel}/PACS & 10 \\
250.0 &   57~$\pm$~10 & \textit{Herschel}/SPIRE & 11 \\
350.0 &   38~$\pm$~15 & \textit{Herschel}/SPIRE & 11 \\
%500.0 & $<~$21.0 & \textit{Herschel}/SPIRE & 11 \\
880.0 &  10.7~$\pm$~5.9 & APEX/LABOCA & 12 \\
1300.0 &  2.0~$\pm$~0.4 & ALMA & 11 \\
9000.0 &0.042~$\pm$~0.014 & ATCA & 13 \\
\enddata
\raggedright

\tablerefs{1. \cite{2015Paunzen}; 2. \cite{1997Mermilliod}; 3. \cite{1997Perryman}; 4. \cite{2003Cutri}; 5. \cite{2010Wright}; 6. \cite{2010Ishihara}; 7. \cite{2009Carpenter}; 8. \cite{2014Chen}; 9. \cite{2001Spangler}; 10. \cite{2012Donaldson}; 11. This work; 12. \cite{2010Nilsson}; 13. \cite{2017Marshall}.}

\end{deluxetable}

\section{Methodology and results}
\label{sec:meth_res}

Here we undertake a determination of the stellar parameters by fitting a grid of stellar atmosphere models to archival high resolution spectra. We also inferred the same stellar parameters as well as the mass and age from stellar evolution models combined with a Bayesian approach. We then present a self-consistent analysis of both the millimetre and scattered light imaging data to determine the disc architecture and dust albedo, respectively. The disc structure is applied as a constraint in modelling of the disc spectral energy distribution (SED) in combination with photometry from archival sources.

\subsection{Stellar parameters}

A summary of the stellar properties used in this work are given in Table \ref{table:star_props}. The stellar position, distance (parallax), proper motions, {and $G$ band magnitude ($G~=~7.3833~\pm~0.0007$)} are taken from the Gaia DR 2 catalogue \citep{2016Prusti,2016Lindegren,2018Brown}. The stellar physical properties were derived by modelling of available high resolution spectra along with optical and near-infrared photometry. 

\begin{deluxetable}{lll}
%\centering
\tablewidth{0.45\textwidth}
\tablecolumns{3}
\tablecaption{{Summary of stellar parameters.} \label{table:star_props}}
\tablehead{
\colhead{Parameter} & \colhead{Value} & \colhead{Ref.} \\
\colhead{}          & \colhead{}      & \colhead{}     \\
}
\startdata
Right Ascension ({\it hms}) & 00 05 52.54      & 1 \\
Declination ({\it dms})     & -41 45 11.0      & 1 \\
Proper motions (mas/yr)     & 97.96, -76.51    & 2 \\
Distance (pc)               & 38.85~$\pm$~0.08 & 2 \\
$V$ (mag)                   & 7.513~$\pm$~0.005 & 3 \\
$B-V$ (mag)                 & 0.600~$\pm$~0.003 & 3 \\
Spectral type               & G0            & 4 \\
Luminosity ($L_{\odot}$)    & {1.216~$\pm$~0.005} & 4 \\
Radius ($R_{\odot}$)        & {1.009~$\pm$~0.003} & 4 \\
Mass ($M_{\odot}$)          & {1.116~$\pm$~0.012}  & 4 \\
Temperature (K)             & {6034~$\pm$~8}    & 4 \\
Surface gravity, $\log g$   & {4.478~$\pm$~0.006} & 4 \\
Metallicity, [Fe/H]         & 0.02~$\pm$~0.04 & 5 \\
$\nu \sin i$ (km/s)         & 17.5~$\pm$~0.5  & 4 \\
Age (Myr)                   & {50~$\pm$~16}   & 4 \\
\enddata
\raggedright

\tablerefs{1. \cite{2007vanL}; 2. \cite{2016Lindegren,2016Prusti,2018Brown}; 3. \cite{1997Mermilliod}; 4. This work; 5. \cite{2014Tsantaki}}
\end{deluxetable}

{The stellar parameters of HD~105, i.e. luminosity, radius, {effective temperature, surface gravity, age}, and mass, were derived by using the absolute $G$ magnitude, parallax, $B-V$ colour, and [Fe/H] as input parameters using the Bayesian approach applied in \cite{2016delBurgo,2018delBurgo}. We derived an age of 50~$\pm$~16 Myr, which is in agreement with that one more constrained assuming membership of Tucana Horologium, of 30 Myr  (del Burgo et al. in preparation). The colour $B-V = 0.600~\pm~0.003$ is taken from \cite{1997Mermilliod}, and {metallicity} [Fe/H]$~=~0.02~\pm~0.04$ from \cite{2014Tsantaki}. The observed $G$ magnitude was converted to absolute magnitude $M_{G}$ using the \textit{Gaia} DR2 parallax of $\pi = 25.75~\pm~0.06$~mas \citep{2018Brown}.} We find that HD~105 is a pre-main-sequence star. {In order to derive the stellar parameters} we downloaded and arranged a grid of PARSEC isochrones \citep[version 1.2S,][]{2012Bressan,2014Chen,2015Chen,2014Tang}, with steps in age of 5 per cent, steps in mass of 0.1 per cent, and steps of {0.02} dex in [Fe/H]. {The lower and upper limits of these parameters are well beyond 6-$\sigma$ around the solution, see Table \ref{table:star_props}. Synthetic photometry estimates were obtained using filter curves from \cite{2018Evans} for the Gaia $G$ band, and \cite{2006MaizApellaniz} for the $B$ and $V$ bands}. For a more detailed description see \citet{2018delBurgo}.

In addition, high resolution FEROS and HARPS spectra were also used to obtain the stellar effective temperature $T_{\rm eff}$, metallicity [Fe/H], projected rotational velocity $\nu \sin i$, and surface gravity $\log g$. We performed a comparison with BTSettl {stellar atmosphere models} \citep{2012Allard}. The models in the grid were modified in order to be compared with the FEROS spectrum similar to the approach of \cite{2009delBurgo}. First, the synthetic spectra were transformed to take into account the stellar projected rotational velocity ($\nu \sin i$) using the formalism of \cite{1992Gray}, with a limb darkening parameter equal to 0.8. These spectra were convolved with a Gaussian that mimics the instrumental profile along the dispersion axis. The resulting spectra were rebinned to the same resolution and grid of the observed spectrum. This was corrected from the velocity shift from a cross-correlation analysis with the models. All modelled and observed spectra were normalized before performing the comparison.

We obtain values from the FEROS spectrum ($R = 48,000$) of $T_{\rm eff} = 6000~\pm~50~$K, $\log g = 4.50~\pm~0.25$, and $\nu \sin i = 17.5~\pm~0.5$~km/s. Values derived from the higher resolution HARPS spectrum ($R = 115,000$) are consistent with those of FEROS, with the FEROS values being determined at higher signal-to-noise. {These values are consistent with those derived from the aforementioned stellar atmosphere models.}

We have a good agreement between the stellar evolution analysis and the stellar atmosphere model fitting for the parameters in common. The values obtained through this analysis are likewise consistent with available values from the literature e.g. \cite{2005ValFis}, \cite{2014Tsantaki}.  

\begin{figure}
\centering
\subfigure{\includegraphics[width=0.5\textwidth]{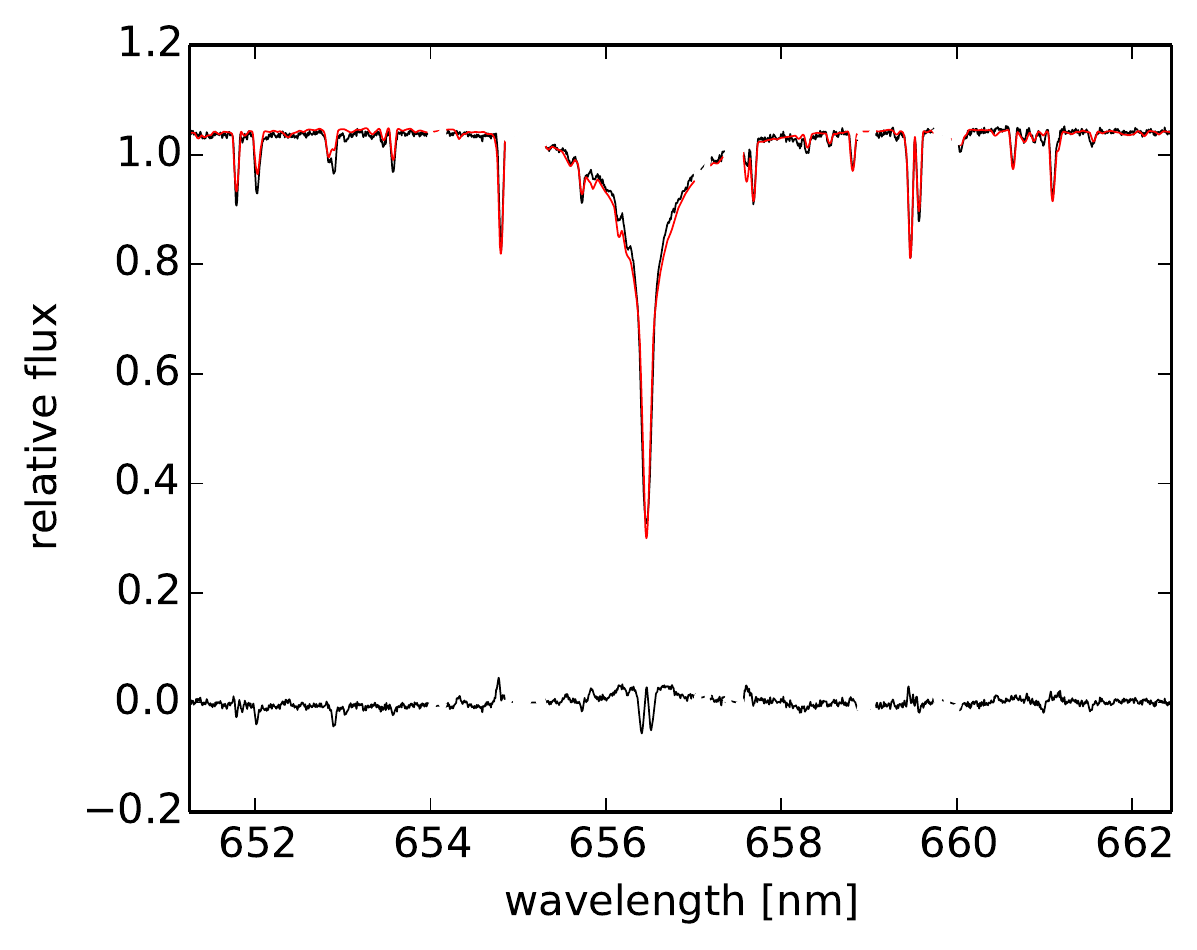}}
\subfigure{\includegraphics[width=0.5\textwidth]{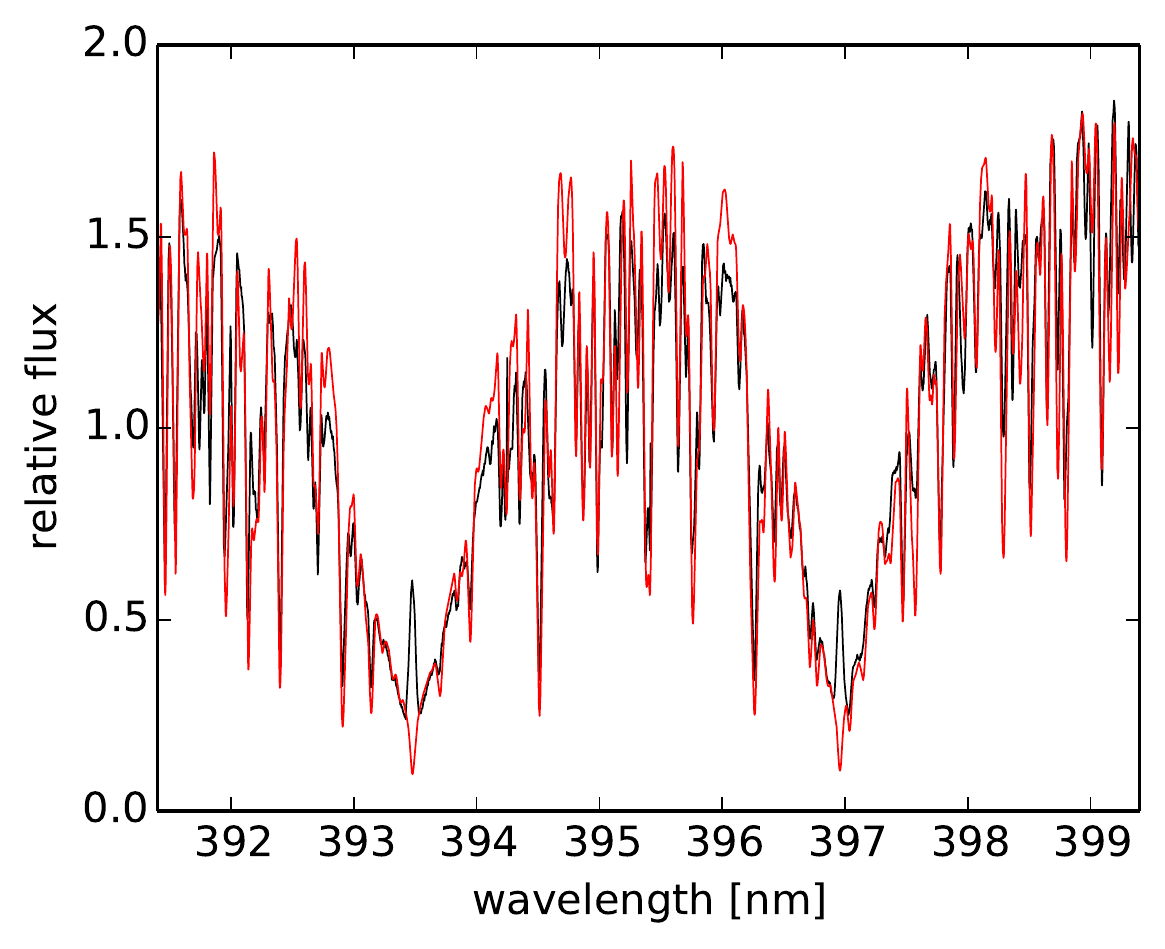}}
\caption{Segments of the FEROS spectrum covering the H$\alpha$ (top) and Ca{\sc ii} H\&K (bottom). The median stellar spectrum from X FEROS observations is presented in black, whilst the best-fitting model is presented in red. Some regions have been blanked out due to contamination from telluric lines. The residual (observed-model) spectrum is shown for clarity in the H$\alpha$ panel. \label{fig:feros}}
\end{figure}

The H$\alpha$ and Ca{\sc ii} H\&K lines in the FEROS spectrum are presented in Fig. \ref{fig:feros}. The median FEROS spectrum, calculated from the available spectra (in black) is plotted together with our best fitting model. Some regions of the H$\alpha$ segment of the spectrum have been blanked out due to contamination from telluric lines at those wavelengths (see Fig. \ref{fig:feros}, top). Despite the low $\nu \sin i$ (consistent with a more pole-on viewing angle), we see clear evidence of Ca{\sc ii} H\&K emission in the spectrum (see Fig. \ref{fig:feros}, bottom). This is consistent with activity in the stellar chromosphere, an indicator of stellar youth. 

We also examined the high resolution spectra for evidence of gas revealed by the presence of circumstellar absorption or emission lines. The origin of such gas may be primordial, or secondary from e.g. photodesorption or cometary activity. We note that the FEROS fibre input has a 2\arcsec\,footprint on the sky that could include light scattered from material in the circumstellar disc, depending on the seeing quality. No evidence of such features is seen. No evidence of gas features in either absorption (implying cool gas) or emission (implying hot gas) were identified in the spectra.

\subsection{ALMA millimetre imaging data}

An ALMA band 6 (1.3~mm) observation for HD~105 was downloaded from the ESO ALMA Science Archive\footnote{\href{http://almascience.eso.org/aq/}{http://almascience.eso.org/aq/}}. The observation was originally carried out as part of project 2012.1.00437.S (PI: D. Rodriguez) during Cycle 1. The spectral setup consists of four windows. Three windows were set up to measure the continuum, each with 128 channels over 2~GHz bandwidth. The fourth covered the $^{12}$CO (2-1) line at 230.538 GHz and sampled its 0.94~GHz bandwith with 3840 channels (0.32 kms$^{-1}$), providing a velocity resolution of 0.64 kms$^{-1}$. In combination, the four channels provide 6.9~GHz bandwidth to study the continuum emission. The on-source integration time for HD~105 was 2238~s. Neptune was used as the flux calibrator, J0006-0623 was the bandpass calibrator, and J0012-3954 was the phase calibrator.

Calibration and reduction of the ALMA observation were carried out in CASA 4.1 using the provided scripts. Image reconstruction was carried out using the \textit{clean} task combining all four spectral windows for the greatest signal-to-noise. We reconstruct the image using natural weighting. With natural weighting the  continuum image r.m.s. noise is 26.8~$\mu$Jy beam$^{-1}$. The dirty beam has an ellipsoidal FWHM $0\farcs95~\times~0\farcs67$ at a position angle of 88$\degr$, equivalent to a spatial resolution of 34$~\times~$26 au.

\subsubsection{Disc architecture}
\label{sssec:resolv}

\begin{figure*}
\centering
\includegraphics[width=\textwidth,trim={2cm 1cm 2cm 1cm },clip]{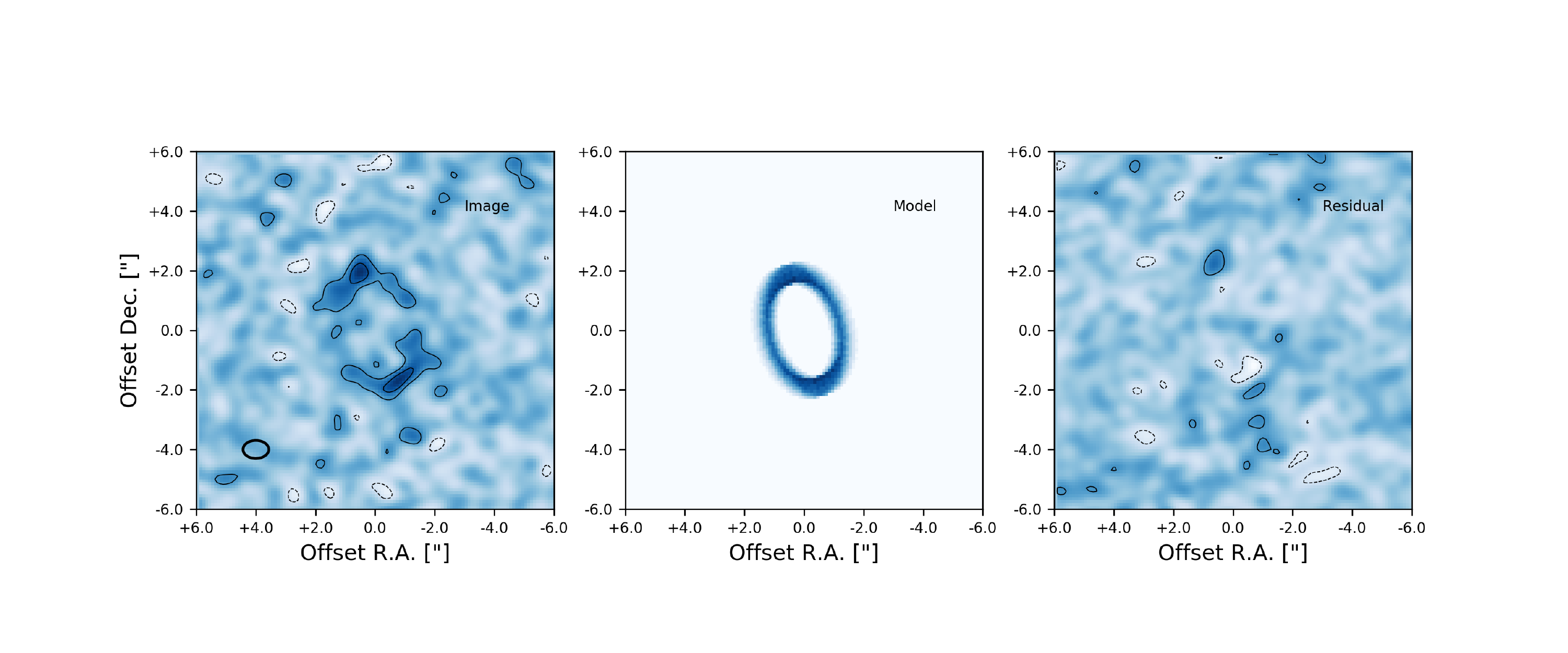}
\caption{ALMA 1.3~mm (Band 6) naturally weighted, dirty image of HD~105, the best fit annular model for the disc, and the residuals in the observed image after subtraction of a model convolved with the dirty beam (from left to right). Orientation is north up, east left. The colour scale for the observations and residuals is linear between -3-$\sigma$ and +3-$\sigma$. The ALMA beam is represented by the black ellipse in the bottom left of the observation. Contours are at -2, 2, 4, and 6-$\sigma$, with the broken contour denoting negative values.\label{fig:hd105_alma}}
\end{figure*}

The disc is visible as an annular structure centred on the stellar position, but the low signal-to-noise of the observations means that it is not detected with high significance ($>~3-\sigma$) at all position angles. The ALMA image and the best fit annular disc model are presented in Fig. \ref{fig:hd105_alma}. 

We model the architecture of HD~105's disc as a single annulus with a semi-major axis $R$ and width $\Delta R$ oriented at a position angle $\theta$ and inclination $i$. The disc surface brightness exponent $\alpha$ is assumed to be constant, and a vertical opening angle of 10$\degr$, similar to the Solar system, was assumed. A grid of models is generated spanning reasonable values for each of these parameters. To determine the best-fit properties for the disc we subtract each disc model, convolved with the dirty ALMA beam (FWHM $\sim~0.9\arcsec\times0.7\arcsec$), from the observed image and seek to minimise the residuals within the region of the model-subtracted image where there is significant emission in the observed image. 

\begin{deluxetable}{lcccc}
%\centering
\tablewidth{0.45\textwidth}
\tablecolumns{5}
\tablecaption{Architecture of HD~105's disc determined from the ALMA image. \label{table:disc_arch}}
\tablehead{
\colhead{Parameter} & \colhead{Range} & \colhead{Values} & \colhead{Spacing} & \colhead{Value} \\
}
\startdata
Inclination ($\degr$)       & 0 -- 90    & 19 & Linear & 50~$\pm$~5 \\
Position Angle ($\degr$)    & 0 -- 90    & 19 & Linear & 15~$\pm$~5 \\
Semi-major axis (au)        & 50 -- 150  & 41 & Linear & 85~$\pm$~5 \\
Semi-major axis ($\arcsec$) &            &    &        & 2.16~$\pm$~0.13 \\
Belt width (au)             & 10 -- 50   & 21 & Linear & 30~$\pm$~10 \\
Belt width ($\arcsec$)      &            &    &        & 0.75~$\pm$~0.25 \\
S.B. exponent, $\alpha$ & 0.0 & 1 & Fixed & 0.0 \\
Total disc flux (mJy)   & 1.0 -- 4.0 & 301 & Linear & 1.45~$\pm$~0.28 \\ 
Dust mass ($M_{\oplus}$)    &            &     &        & 0.035~$^{+0.013}_{-0.009}$ \\
\enddata
\raggedright
\end{deluxetable}

The best-fit architecture derived from the model fitting is dominated by the disc ansae, which are the only regions of the disc detected at high signal-to-noise ($>$~3-$\sigma$). The disc semi-major axis is 2\farcs16~$\pm$~0\farcs13 (85~$\pm$~5~au), {with} an inclination of 50~$\pm$~5~\degr with respect to the line of sight at a position angle of 15~$\pm$~5\degr. The width of the disc annulus is 0\farcs75~$\pm$~0\farcs25 (30~$\pm$~10~au). The belt is perhaps marginally resolved in the ALMA image, but the quality of the data does not allow us to conclude that, so we hereafter assume the belt is unresolved. The best-fit disc architecture obtained from this approach is summarised in Table \ref{table:disc_arch}. 

The total flux of the disc in the sub-millimetre, 1.45~$\pm$~0.28~mJy, may be used to calculate the mass of dust grains, assuming that the disc is fully optically thin at 1.3~mm. For a disc observed at a frequency $\nu$ with a flux $F_{\nu}$ the dust mass, $M_{\rm dust}$, is given by
\begin{equation}
	M_{\rm dust} = F_{\nu} d_{\star}^{2} / \kappa_{\nu} B_{\nu, T_{\rm dust}}
\end{equation}
where $d_{\star}$ is the stellar distance, $\kappa_{\nu}$ is the dust opacity, $T_{\rm dust}$ is the dust temperature, and $B_{\nu,T_{\rm dust}}$ is the Planck function \citep{1993ZucBec,2006Draine}. Assuming the dust opacity $\kappa$ is 1.7~gcm$^{-2}$ \citep{1990Beckwith,1994Pollack,2006Draine}, that is commonly assumed for debris dust \citep[e.g. ][]{2013Panic,2017Holland}, we calculate a mass of $0.035^{+0.013}_{-0.009}~$M$_{\oplus}$ for the dust. This value does not include uncertainty on the parameter $\kappa$, which leads to an uncertainty in $M_{\rm dust}$ of a factor 3 to 5.

\subsubsection{Non-detection of cold CO emission}
\label{sssec:no_gas}

Molecular carbon monoxide (CO) has been identified in around a dozen debris disc systems \citep[e.g.][]{2016Greaves,2017Moor}, believed in most cases to be the result of secondary production after liberation from exocomets \citep{2017Kral}. Most of the stars identified with gas in their debris discs are young ($t_{\rm age}~<~30$~Myr) A-type stars, with relatively bright debris discs ($L_{d}/L_{\star}~\sim~10^{-3}$). Here we search the ALMA observation for emission from the disc associated with the CO (2-1) transition at 230.538~GHz following the method employed to extract CO emission at low signal-to-noise in ALMA data as presented in \cite{2015Matra,2016Marino,2017Marino}. 

To begin we produced a continuum subtracted measurement set from the HD~105 data using \textit{uvcontsub}. The continuum subtracted data set is then \textit{clean}ed to produce an image cube with the same weighting and pixel scale as the continuum image. The cube spans barycentric velocities between -25 and 25~kms$^{-1}$, with a step size of 0.5~kms$^{-1}$. The cube is corrected for the effect of the primary beam using the \textit{impbcor} task before further analysis. No significant CO emission is observed in the velocity-integrated, continuum-subtracted image produced by integrating the cube.

We then proceed by implementing the spectro-spatial filtering technique presented in \cite{2017Matra}. Any circumstellar CO is assumed to originate from collisions between planetesimals and therefore reside in the same region around the star as the debris dust detected in the continuum image. The CO is further assumed to be situated in a vertically flat disc, orbiting with Keplerian velocity around a star of 1.114~$M_{\odot}$. We then determine the projected radial velocity for each position in the disc image given the extent and inclination of the disc for cases of the disc rotating toward or away from us. Using the predicted velocity field the spectrum at each position is then shifted to match the stellar velocity (i.e. spectral filtering). We then integrate over all positions where continuum emission is detected to avoid contributions to the CO spectrum from locations and velocities where no circumstellar CO is expected (i.e. spatial filtering). The spectro-spatial filtering yields a 3-$\sigma$ upper limit of 8.5~mJy kms$^{-1}$~to CO emission. 

We use a {non-Local Thermodynamic Equilibrium} CO excitation code to calculate a CO mass upper limit, following \cite{2018aMatra}. This code includes fluorescent excitation from the central star at ultraviolet wavelengths. Depending on the excitation conditions we determine a CO mass upper limit ranging from 2.5$\times10^{-7}$ to 2.5$\times10^{-6}$~$M_{\oplus}$. This translates to upper limits on the CO and CO$_{2}$ fraction (by mass) in exocomets between $<~88~\%$ and $<~46~\%$ for the respective mass upper limits. This limit is consistent with the range of values for other exocometary discs, and Solar system bodies \cite[see e.g.][]{2017Matra}.

\subsection{Detection of scattered light}
\label{sssec:sca}

HD~105 has been observed by both \textit{HST}/NICMOS and VLT/SPHERE. The disc was not directly visible in either data set. However, we recover the disc at the correct angular separation and orientation using an angular averaging technique to aggregate the disc emission to a detectable level. From this measurement of the disc scattered light brightness we determine the albedo of the dust grains at near-infrared wavelengths.

\subsubsection{HST/NICMOS}

HD~105 was observed with the \emph{Hubble Space Telescope} (HST) NICMOS instrument at two epochs using the coronagraphic mode (mask radius 0\farcs3) of the mid-resolution NIC2 channel (0\farcs07565~pixel$^{-1}$). The first data were acquired on UT-1998-11-03 as part of program GTO-7226 (PI: E. Becklin), a survey looking for giant planets around young nearby stars \citep{2005Lowrance}, using the F160W filter of bandpass very comparable to the H band (pivot wavelength $1.600~\mu$m, 98\%-integrated bandwidth $0.410~\mu$m). The second dataset was obtained on UT-2006-07-23 as part of program GO-10527 (PI: D. Hines), a survey looking for debris discs around stars with infrared excess detected with \emph{Spitzer} as part of the FEPS program \citep{2008Hillenbrand}. The data were obtained with the F110W filter, which is twice as extended toward shorter wavelengths as the J band (pivot wavelength $1.116~\mu$m, 98\%-integrated bandwidth $0.584~\mu$m).

The F160W data were acquired at two different spacecraft orientations, obtaining a total of 6 exposures with HST rolled by $\sim~30\degr$ in the middle of the sequence, to enable subtraction of the PSF with Roll Differential Imaging \citep{1999Lowrance}. The total exposure time is 1344~s for this dataset. The F110W dataset includes 7 exposures all obtained with the same telescope orientation, for a total exposure time of 2016~s.

We reprocessed both datasets as part of the \emph{Archival Legacy Investigations for Circumstellar Environments} (ALICE) program (PI: R. Soummer), a consistent re-analysis of the NICMOS coronagraphic archive using modern PSF subtraction techniques \citep{2014dChoquet,2018Hagan}. Assembling PSF libraries from multiple reference stars in the whole archive and using the PCA-based KLIP algorithm for PSF subtraction \citep{2012Soummer}, this program demonstrated a gain in point source sensitivity by a factor of 20 at $1\arcsec$ from the star over classical reference star differential imaging \citep{2016Debes}, and enabled the detection of 11 faint debris discs from NICMOS archival data \citep{2014Soummer,2016Choquet,2017Choquet,2018Choquet}. 

\begin{figure*}
\centering
\includegraphics[width=0.33\textwidth,trim=-2cm 0cm 0cm -1cm, clip]{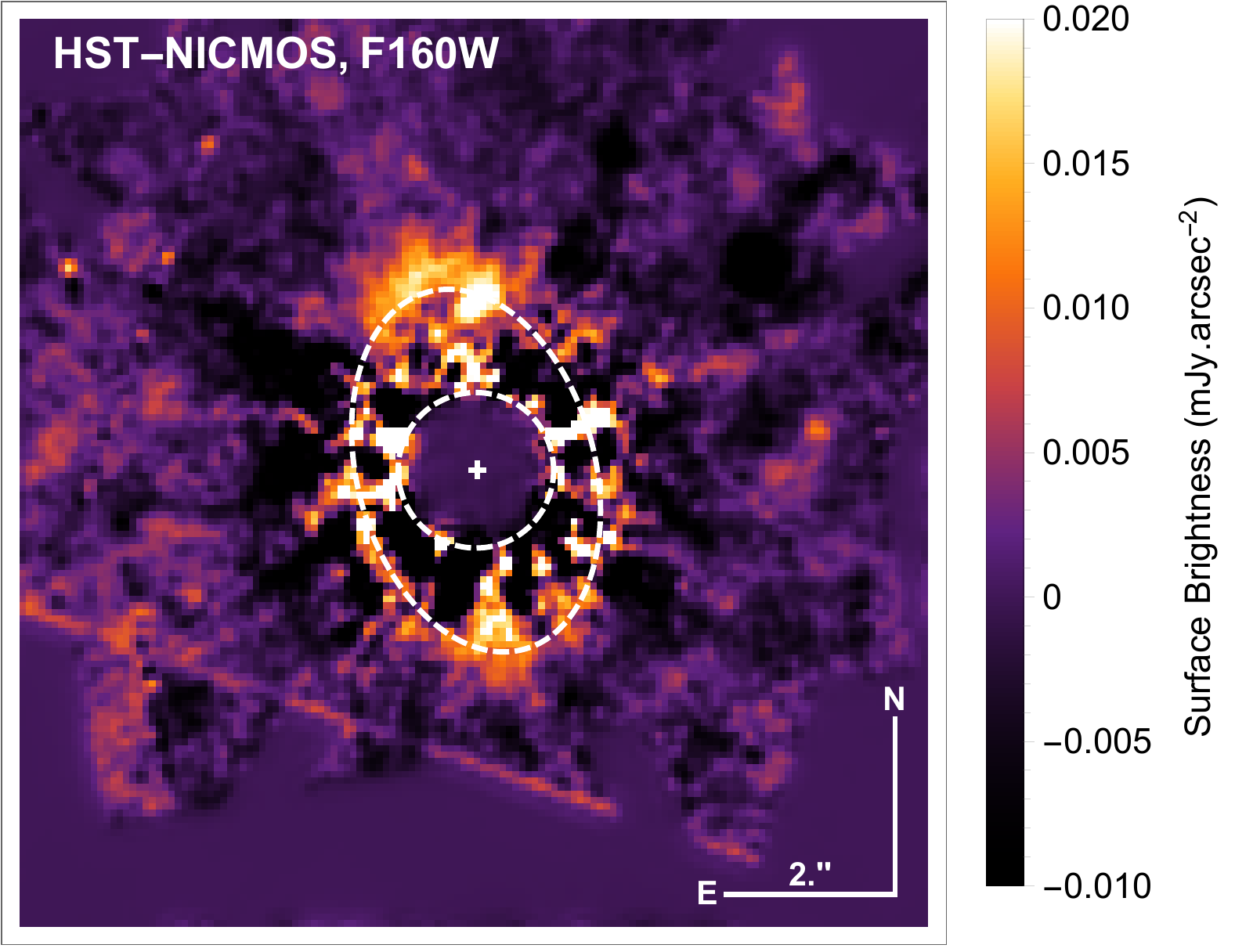}
\includegraphics[width=0.33\textwidth,trim=-2.6cm 0cm 0cm -1cm, clip]{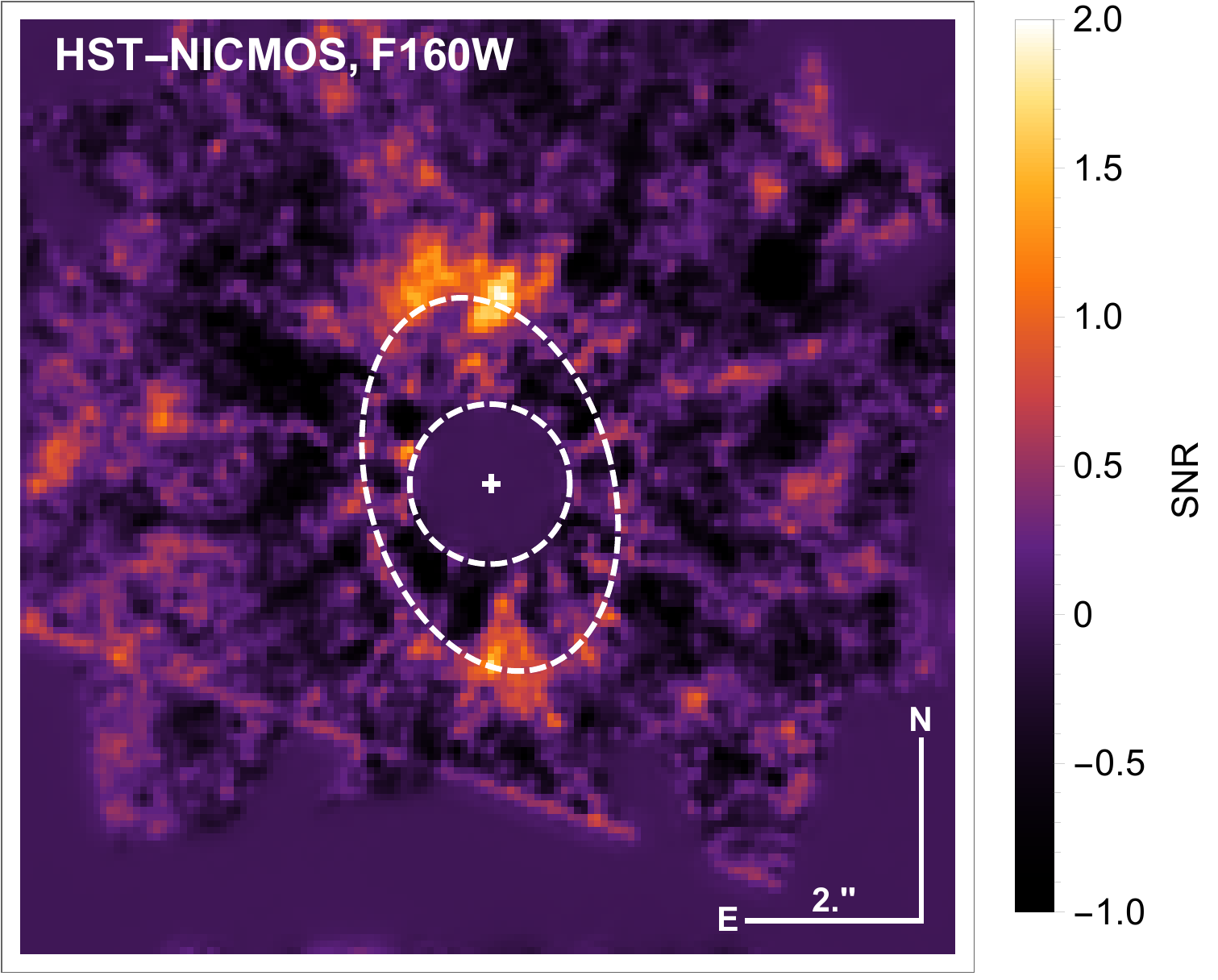}
\includegraphics[width=0.33\textwidth]{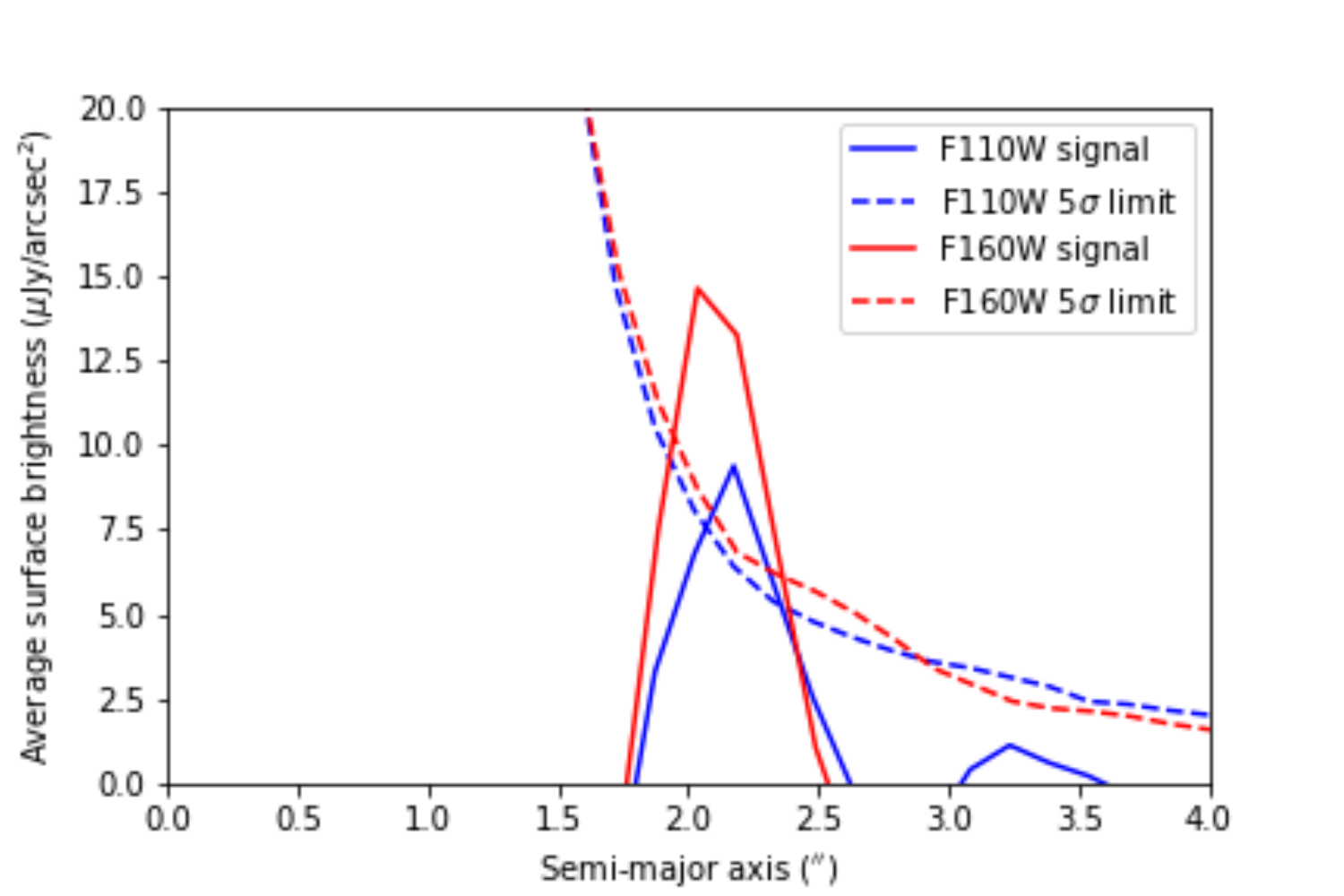}
\caption{\textit{Left}: \textit{HST}/NICMOS F160W image of HD~105, smoothed by convolution with a model PSF. Significant emission from the disc is recovered at the ansae. The `+' symbol denotes the stellar position, the circular dashed line denotes the {numerical} mask, and the dashed elliptical line denotes the extent and orientation of the debris disc from the ALMA image. Orientation is north up, east left. {\textit{Middle}: Signal-to-noise map of the \textit{HST}/NICMOS image also smoothed by convolution with a model PSF. The noise was estimated from from the pixel-wise standard deviation of reference stars PSF-subtracted with the same parameters as for HD 105 \citep{2018Choquet}.} \textit{Right}: Results of {elliptical} averaging of the \textit{HST}/NICMOS F110W and F160W images, revealing significant emission at the orientation and semi-major axis of HD~105's disc as imaged by ALMA ($i = 50\degr$, $\theta = 15\degr$). The signal peaks in the F110W filter around 10$\mu$Jy/arcsec$^{2}$ at a separation of 2.2\arcsec, and the signal in the F160W filter peaks at 15$\mu$Jy/arcsec$^{2}$ at a similar separation. The solid blue (red) line denotes the extended emission in F110W (F160W) filter, whilst the dashed red (blue) line denotes the 5-$\sigma$ surface brightness limit in the respective filters. \label{fig:hst_scattered}}
\end{figure*}

For HD~105, we used the reprocessed data from the ALICE public database\footnote{\href{https://archive.stsci.edu/prepds/alice/}{https://archive.stsci.edu/prepds/alice/}} for the F110W dataset, and an unpublished ALICE reduction for the F160W dataset with an $11\times11\arcsec$ field of view, more favorable to disc detection with sensitivity limits better than the data published in the ALICE database. PSF subtraction for the F160W dataset was achieved using the 69 first eigen-modes of a library assembling 125 images from reference stars exclusively, and excluded a central zone within a radius of 12 pixels. The F110W dataset was PSF-subtracted with the 211 first eigen-modes of a 640-image library from reference stars only, excluding a central area of radius 5 pixels. The final images were obtained by rotating all exposures to North pointing up, co-adding and scaling them to surface brightness units based on NIC2's plate-scale and its calibrated photometric factors ($F_{\nu}=1.21121~\mu$Jy s DN$^{-1}$ in F110W, $F_{\nu}=1.49585~\mu$Jy s DN$^{-1}$ in F160W). We are able to recover the disc scattered light in the F160W filter image {in the disc's ansae at low signal-to-noise}, as shown in Fig. \ref{fig:hst_scattered}.

To obtain a {more significant} measurement of the disc surface brightness than can be obtained from the reduced images, we make use of the high spatial resolution of the scattered light image and measure the surface brightness radial profiles. {We computed the surface brightness averaged in 5-pixel wide ($0.38\arcsec$) elliptical apertures of increasing semi-major axes $a$ with semi major axis $a$ increasing with steps of 1 pixel (76~mas) and semi-minor axis $b = a \cos(50\degr)$, oriented $15\degr$ east-of-north, to trace the disc deprojected average radial profile according to the ALMA detection geometry}. We estimate the corresponding uncertainties assuming uncorrelated noise, as $\sigma/\sqrt(N_{\rm pix})$ with $\sigma$ the standard-deviation in each aperture and $N_{\rm pix}$ the number of pixel in the aperture. Since only reference stars were used for PSF subtraction, these measurements are not affected by disc self-subtraction. They are however subject to over-subtraction by the PCA algorithm along with PSF features \citep{2012Soummer,2016Pueyo}. We corrected our {average profiles and uncertainties} by estimating the algorithm throughput with analytical forward modeling. From this exercise we find a significant peak with a surface brightness of 15~$\pm$~1~$\mu$Jy/arcsec$^2$ at a separation of around 2.2\arcsec\, from the star, consistent with the ALMA image. The binned radial profiles are presented in Fig. \ref{fig:hst_scattered}.

\subsubsection{VLT/SPHERE}

HD~105 was observed with the Spectro-Polarimeter High-contrast Exoplanet REsearch \cite[SPHERE,][]{2008Beuzit} on the night of October, 2nd 2015. These observations used the Infra-Red Dual-beam Imager and Spectrograph \cite[IRDIS,][]{2008Dohlen} of SPHERE to obtain high-contrast and high-angular resolution images of the circumstellar environment of HD~105, in the broad-band H filter (centred at 1.625~\micron, width 291~nm) with the apodized Lyot coronagraph of diameter 185mas. They are part of the SPHERE High-Angular Resolution Debris Disc Survey\footnote{ESO program ID 096.C-0388(A)} \citep[SHARDDS,][]{2018Milli}. SHARDDS is an open-time program on SPHERE to perform the first comprehensive near-infrared survey of all bright, nearby debris discs (infrared excess greater than $10^{-4}$), yet undetected in scattered light, within 100~pc. It already led to {detections of several discs and a brown dwarf companion}: HD~114082 \citep{2016Wahhaj}, 49 Ceti \citep{2017Choquet} and HD~206893 \citep{2017Milli}. HD~105 was observed for a total of 50~min on-source, among which the total exposure time is 2432~s, in pupil-stabilized mode to allow Angular Differential Imaging \citep{2006Marois}. The source was observed just before meridian crossing, which resulted in a parallactic angle rotation of $35\degr$.

The data were sky-subtracted, flat-fielded and bad pixel corrected using the official SPHERE Data Reduction and Handling pipeline \citep{2008Pavlov} in order to make a temporal cube of 606 frames. Each coronagraphic frame was re-centred using the set of four satellite spots imprinted in the image during the centring sequence. This sequence is obtained by applying a waffle pattern to the deformable mirror and was done prior to and after the deep science coronagraphic observations. Then, different data reduction techniques were applied to subtract the stellar halo and try to detect the scattered light of the disc. 

At the expected semi-major axis of the disc of 2.16\arcsec, the technique that yields the best sensitivity for point source detection is the classical Angular Differential Imaging \citep[cADI,][]{2006Marois} technique. However for an azimuthally extended structure such as the HD~105 disc, the throughput of the algorithm is very low because of self-subtraction \citep{2012Milli} and we estimated it to be a few percent. The 5-$\sigma$ sensitivity of the ADI reduction is 100$\mu$Jy at 2.2\arcsec after throughput correction, and we are therefore unable to confirm the \textit{HST}/NICMOS detection with this data reduction technique. 

Therefore, we also applied the Reference Star Differential Imaging (R[S]DI) technique, commonly used in space-based observations \citep{1999Schneider,2009Lafreniere,2011Soummer}, and more recently applied to ground-based observations \citep{2016Gerard}. We implemented this technique in the SHARDDS program by building a library of all the coronagraphic images obtained in the program. We then selected 610 images from 22 stars most correlated with the ones from HD~105 (excluding the HD~105 images themselves) and used this set of references to subtract the stellar halo, using a custom PCA algorithm, keeping 100 modes. The disc is not directly visible in the VLT/SPHERE image.

{The image is binned by a factor 4, giving a new pixel size of 49~mas, corresponding to the measured FWHM in the non-coronagraphic image. In the image reduced with PCA-RDI, the flux is integrated in elliptical apertures having an aspect ratio matching the inclination of the disc as seen by ALMA, and a radial width of 10 binned pixels or 490~mas. This ellipse is increased iteratively by step of 1 binned pixels or 49~mas, and at each iteration, the signal is computed as the sum of the flux over this elliptical aperture. It is corrected for the PCA throughput and converted in $\mu$Jy/arcsec$^{2}$ using the binned pixel size of 49~mas and the star flux as measured in the non coronagraphic image. The noise is given by the standard deviation of the pixels over the same elliptical annulus.}

\begin{figure*}
\centering
\includegraphics[width=0.33\textwidth,trim=2cm 0cm 0cm 1cm, clip]{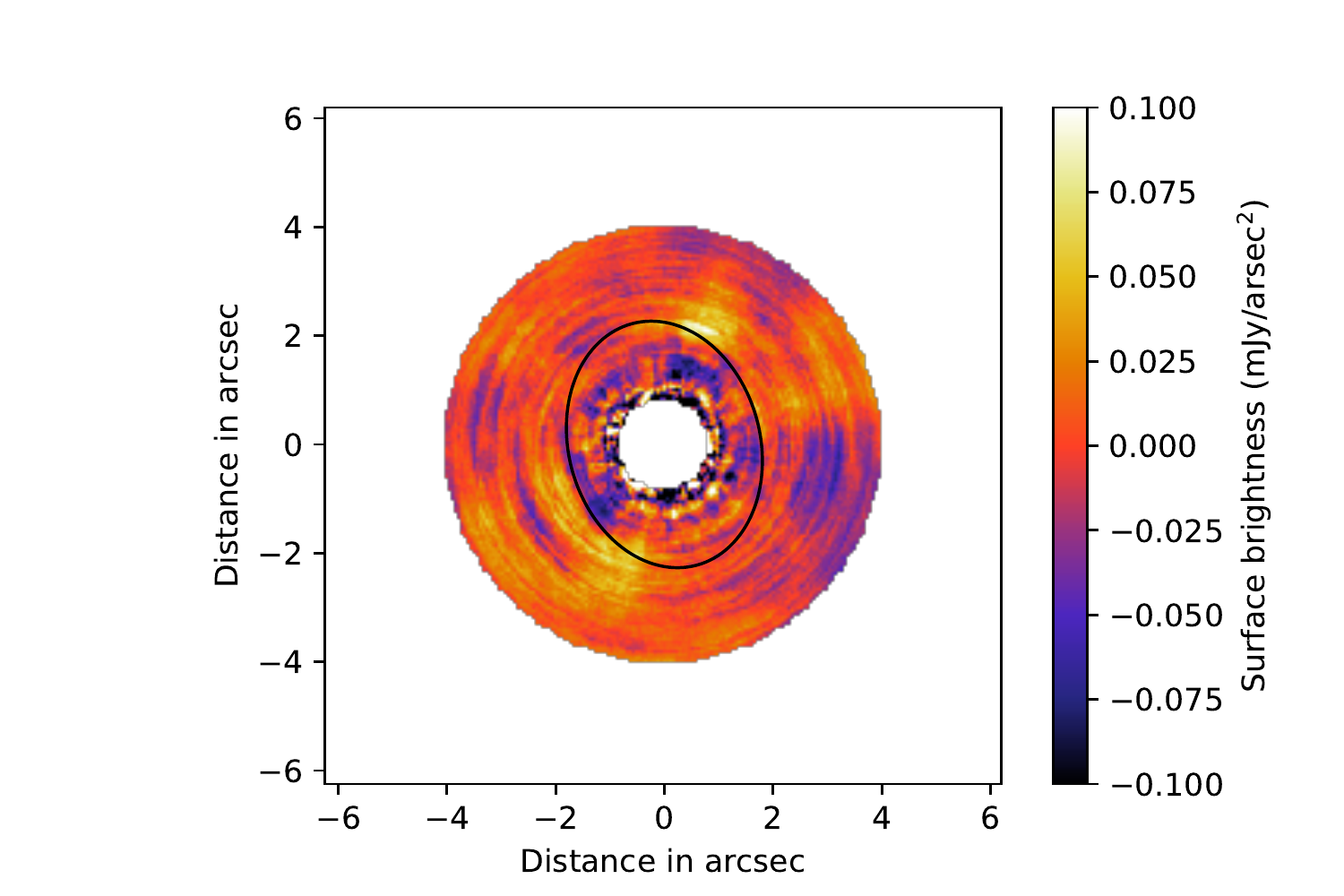}
\includegraphics[width=0.33\textwidth,trim=2cm 0cm 0cm 1cm, clip]{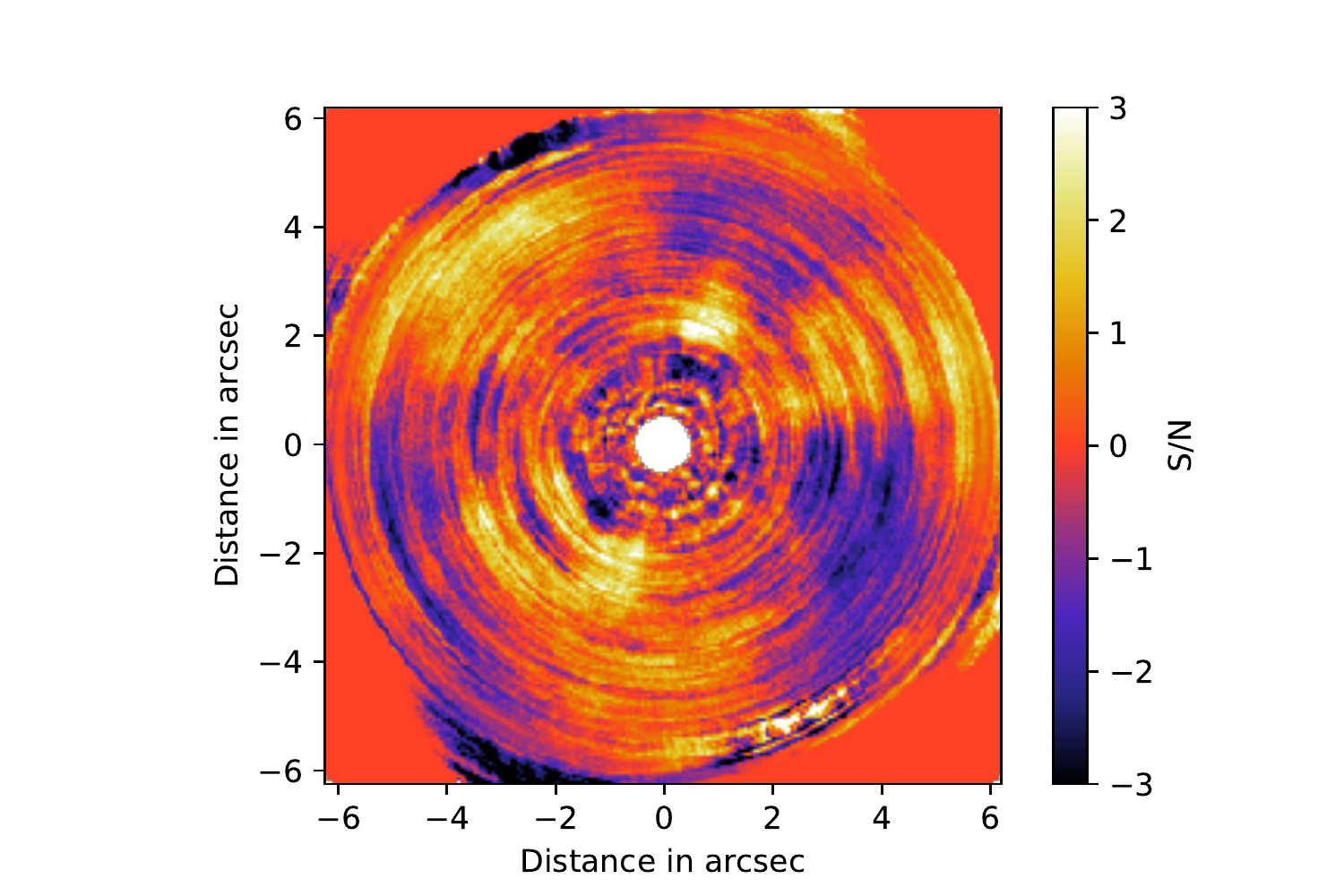}
\includegraphics[width=0.33\textwidth]{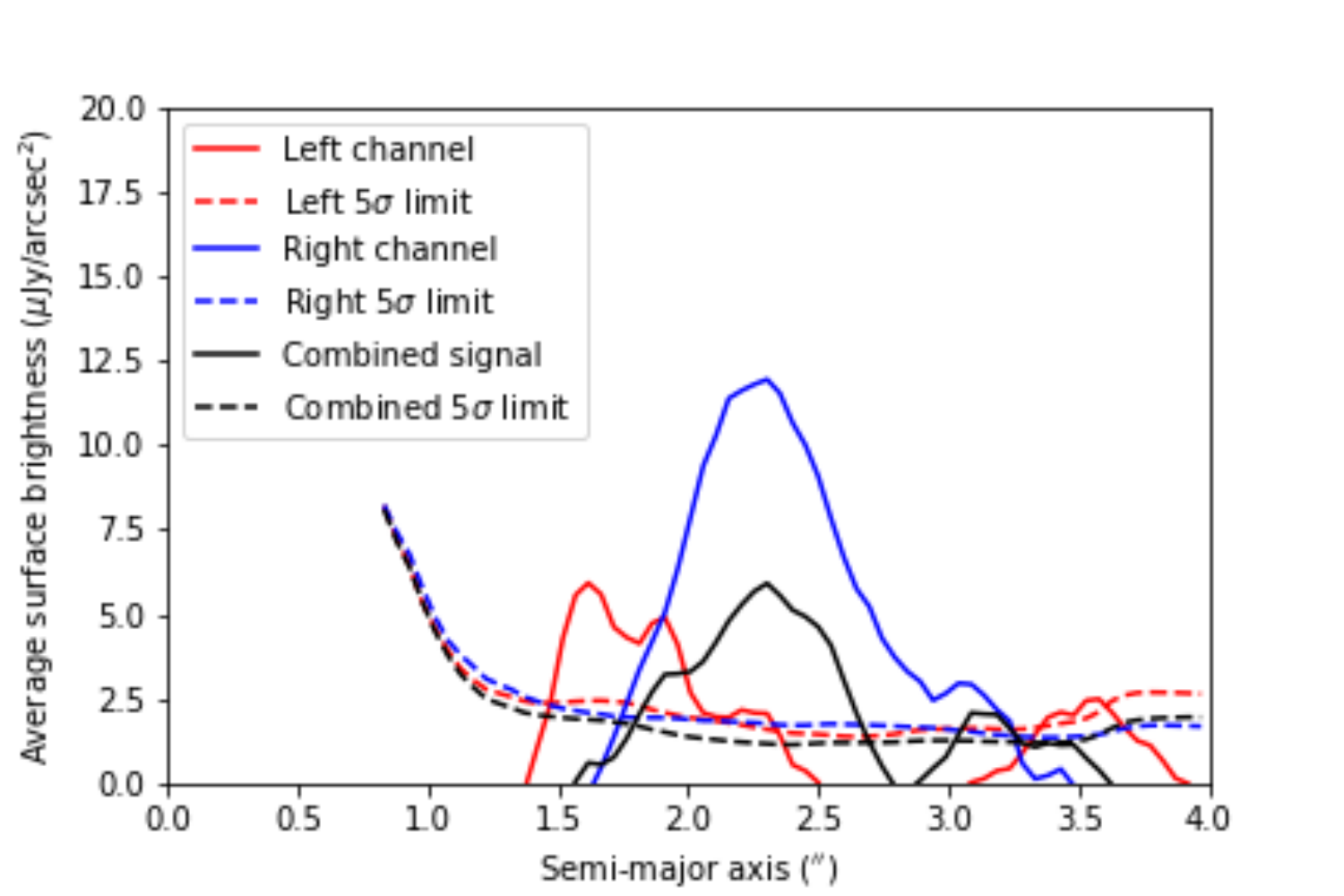}
\caption{\textit{Left}: VLT/SPHERE PCA-RDI reduced image of HD~105. The black ellipse denotes the location of the disc in the ALMA image. {\textit{Middle}: Signal-to-noise map of the VLT/SPHERE image, combining both left and right detector channels.} \textit{Right}: Results of elliptical annular averaging of the VLT/SPHERE image, revealing significant emission at the orientation and semi-major axis of HD~105's disc as imaged by ALMA ($i = 50\degr$, $\theta = 15\degr$). The emission peaks at 5.9~$\mu$Jy/arcsec$^{2}$ at a separation of 2.3\arcsec, denoted by the black solid line in the figure; the black dashed line denotes the 5-$\sigma$ surface brightness limit. The red and blue lines denote the measured brightness and 5-$\sigma$ limits for the left and right channels, respectively. We note that the two channels are very different, but are consistent in they reveal significant emission between 1.5 to 3.0$\arcsec$ from the star on the same orientation as the disc in the ALMA image. See text for details. \label{fig:sphere_scattered}}
\end{figure*}

{This averaging technique increases the depth to which we can detect disc scattered light emission using geometrical prior from the resolved mm detection.} We find a peak in emission at 5.9~$\mu$Jy/arcsec$^{2}$ for a semi-major axis of 2.3\arcsec, consistent with the disc semi-major axis, as shown in Fig. \ref{fig:sphere_scattered}. This is a factor of $\sim$4 fainter than the disc detected by \textit{HST}/NICMOS; examination of the separate channels reveals that the right channel is consistent with the NICMOS-detected signal in both semi-major axis and brightness, but the left channel is not. Both channels are consistent in the presence of emission between 1.5 and 3.5\arcsec, consistent with HD~105's disc. The origin of the discrepancy between the \textit{HST}/NICMOS and VLT/SPHERE detections potentially lies in the estimation of noise, which has been assumed Gaussian here, but it can be seen in the reduced image that the noise is non-Gaussian.

\subsubsection{Combined modelling}

The scattered light measurements from \textit{HST}/NICMOS and VLT/SPHERE can be used to calculate the albedo of the dust grains. Assuming optically thin dust, and using the approximation from \cite{1999Weinberger}, the maximum allowed optical depth $\tau$ times the albedo $\omega$ is $4\pi\phi^2 S/F$ where $S$ is the surface brightness of the disc in mJy/arcsec$^2$, $F$ the total star flux in mJy and $\phi$ the separation of the scatterers. Approximating the optical depth by $\tau = 2 f \phi \cos(i) / [ d\phi (1-\omega) ]$ where $d\phi$ is the disc width in arcsec and $f$ the fractional luminosity, we obtain:
\begin{equation}
S = \frac{f F \omega}{(2\pi \times \phi \times d\phi \times \cos(i) )(1-\omega)}
\end{equation}
Using a disc semi-major axis $\phi=2.16\arcsec$ with a width $d\phi=0.75\arcsec$ at an inclination $i=50\degr$ we obtain an albedo of 0.15 based on the NICMOS surface brightness (15~$\mu$Jy/arcsec$^{2}$), or 0.06 based on the SPHERE measurement (5.9~$\mu$Jy/arcsec$^{2}$). The apparent surface brightness can be significantly enhanced in case of strong forward scattering \citep{2015HedSta,Milli2017_hr4796,2016Olofsson} but given the inclination of the HD~105 disc scattering angles below $40\degr$ are not probed. 

With the radial averaging, we have been able to measure dust albedoes a factor of 3 to 10 deeper than expected. The detection of HD~105 in scattered light made here points the way to recovery of similarly faint discs in scattered light where the disc geometry is already known from spatially resolved continuum imaging. Deeper near-infrared or optical images are required to spatially resolve the scattered light disc, in particular to determine the degree of asymmetry in the dust scattering.

\subsection{{Constraints on the dust composition}}

{Using the mean surface brightness from the scattered light detections we can use the inferred albedoes to constrain the dust grain composition. We calculate the scattering albedo $\omega = F_{\rm scat}/(F_{\rm scat}+F_{\rm therm})$, where $F_{\rm scat}$ is the scattered light fractional luminosity in the F160W filter, and $F_{\rm therm}$ is the dust continuum fractional luminosity. Following \cite{2018Choquet}, we assume that the debris disc to be composed entirely of dust grains with a single size and of a single composition. It is further assumed that the same dust grains responsible for the scattered light are also responsible for the thermal emission so the value of $F_{\rm therm}$ is therefore held fixed for all the tested combinations of grain size and compositions, i.e. $L_{\rm dust}/L_{\star} = 2.6\times10^{-4}$. For consistency, we keep the dust mass of the disc fixed to that calculated for HD~105 based on its ALMA mm flux.}

{For each combination of grain size and composition we calculate the scattered light brightness of the disc at the observed orientation and inclination of HD~105, convolving the disc's scattered light SED with the F160W bandpass to determine the scattered light fractional brightness $F_{\rm scat}$, and hence the scattering albedo $\omega$. The scattered light modelling is carried out using the 3-D radiative transfer code {\sc Hyperion} \citep{2011Robitaille}. We test grain sizes from 0.1 to 500~$\mu$m, and dust compositions of pure astronomical silicate, and water ice:astronomical silicate mixtures in the ratios 10:90, 30:70, and 50:50. Optical constants for the materials were taken from \cite{2003Draine} (astronomical silicate) and \cite{1998LiGre} (water ice). Calculation of the optical constants for composite materials was done using the Bruggeman effecive medium theory \citep{1935Bruggeman}.}

{We find that dust grain sizes of a few microns, and either pure silicate or moderately icy compositions, are consistent with the dust albedo derived from the NICMOS-observed mean disc surface brightness.}

\begin{figure}
\centering
\includegraphics[width=0.5\textwidth]{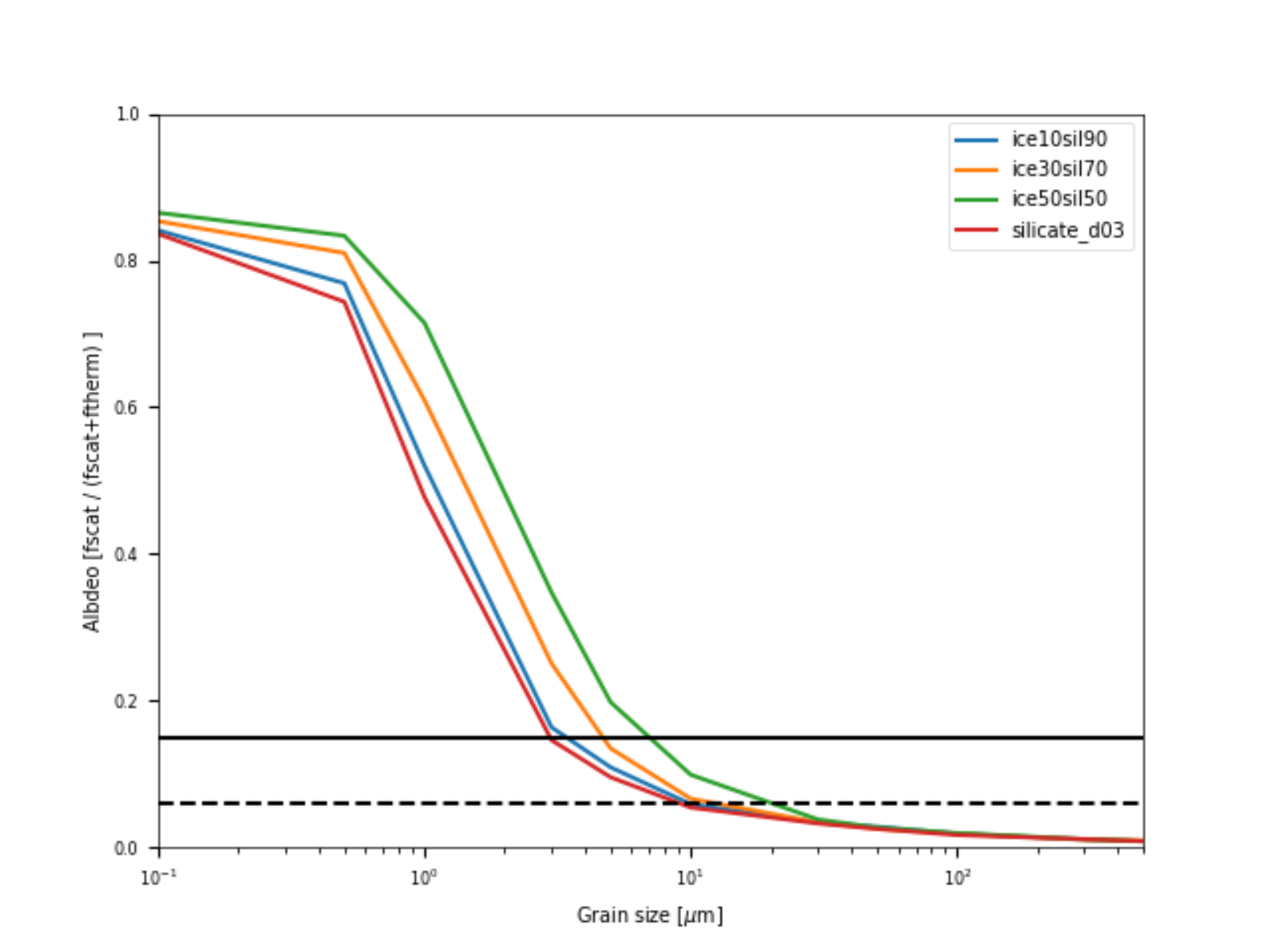}
\caption{{Plot of dust grain size vs. scattering albedo illustrating the constraints on dust grain size and composition possible with the new scattered light detection of HD~105's disc. The horizontal black lines denote the expected dust albedo required to reproduce the mean surface brightness from NICMOS (solid, $\omega = 0.15$) and SPHERE (dashed, $\omega = 0.06$). The coloured lines denote the scattering albedoes for discs composed of dust grains with a single size and composition. We find that silicate (or slightly icy) dust grains around a few $\mu$m in size are consistent with the NICMOS-derived scattering albedo; larger grains ($\geq~10~\mu$m) are required to reproduce the SPHERE-derived albedo, but this would be inconsistent with the typical grain size of dust around Sun-like stars of a few to ten times the blow-out radius.} \label{fig:dust_alb}}
\end{figure}

\subsection{Revised SED model}

Here we combine the results previously determined from each dataset in order to model the full system. We use the disc architecture, as determined from the ALMA image, {and the dust scattering albedo measurement from the scattered light images, as constraints in a radiative transfer model of the debris disc}. We model the disc as being an annulus lying 70 to 100 au from the star with uniform radial surface brightness, as per the ALMA image. The disc is composed of {compact, spherical} dust grains described by a power law size distribution ($dn \propto a^{-q} da$) with exponent $q$ spanning a size range from $a_{\rm min}$ to $a_{\rm max}$. The minimum dust grain size is a free parameter, but the maximum grain size is fixed at 10~mm to best replicate the millimetre wavelength photometry. {The dust composition is assumed to be pure astronomical silicate \citep{2003Draine}.}

\begin{figure}
\centering
\includegraphics[width=0.5\textwidth]{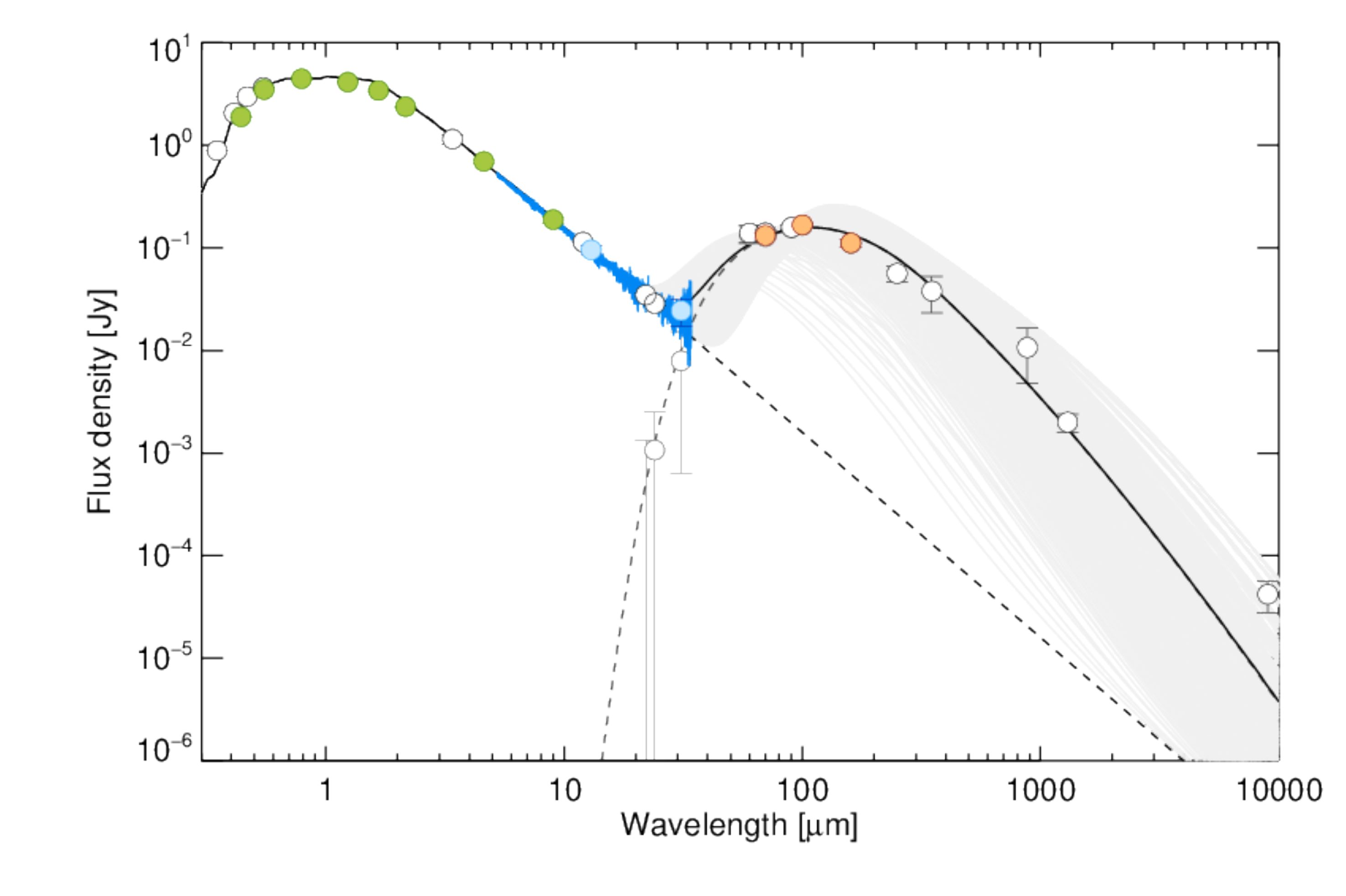}
\caption{SED of HD~105, from power law modelling. Green dots denote optical and infrared photometry used to scale the stellar photosphere model. Blue lines and dots denote \textit{Spitzer}/IRS data. Orange dots denote \textit{Herschel}/PACS data. White dots denote ancilliary photometry from various literature sources (see Table \ref{table:disc_phot}). Uncertainties are 1$\sigma$. Dark grey data points denote excess values. The greyed-out region denotes the envelope of disc models produced from the grid parameter space. The dashed line denotes the individual stellar photosphere and disc components, whilst the solid line is the total emission.\label{fig:sed_bf}}
\end{figure}

The resultant disc SED based on the best-fit model parameters is presented in Fig \ref{fig:sed_bf}, and a summary of the best-fit disc properties (including the disc extent as fixed input) is provided in Table \ref{table:disc_pwrlaw}. The stellar photosphere component (dashed black line) is represented by a Castelli-Kurucz model \citep{2004CK} of appropriate to the stellar spectral type ($\log g = 4.5$~cm/s$^{2}$, $T_{\rm eff} = 6000$~K, [Fe/H] = 0.02) scaled to the optical and near-infrared fluxes (green data points). Blue and red data points denote \textit{Spitzer}/IRS and \textit{Herschel}/PACS fluxes, respectively. Grey data points are the excess (i.e. total - star) fluxes for wavelengths $>~20\mu$m. The individual models comprising the model grid used fit the observations is represented by the set of solid light grey lines, with the best-fit disc model denoted by the dashed dark grey line. The total model of the system (star+disc) is denoted by the solid black line.

\begin{deluxetable}{lcccc}
%\centering
\tablewidth{0.45\textwidth}
\tablecolumns{5}
\tablecaption{Modelling results from power law fit to disc emission. \label{table:disc_pwrlaw}}
\tablehead{
\colhead{Parameter} & \colhead{Range} & \colhead{Values} & \colhead{Spacing} & \colhead{Value} \\
}
\startdata
Semi-major axis (au)        &     & 1 & Fixed & 85  \\
Belt width (au)             &     & 1 & Fixed & 30  \\
$\alpha$                    & 0.0 & 1 & Fixed & 0.0 \\
\hline
$a_{\rm min}$ ($\mu$m) & 0.5 -- 20.5 & 41 & Linear & 3.25$^{+3.00}_{-1.00}$ \\
$a_{\rm max}$ ($\mu$m) & 10$^{4}$ & 1 & Fixed & 10$^{4}$ \\
$q$                    & 2.50 -- 4.50 & 11 & Linear & 3.55$^{+0.30}_{-0.15}$\\
\hline
Composition & --- & 1 & --- & Astro. sil. \\
\enddata
\raggedright
\end{deluxetable}

\section{Discussion}
\label{sec:dis}

There is a growing body of work examining the resolved extent of debris discs as a function of their host stars' luminosities \citep[e.g.,][]{2013Booth,2013Morales,2014Pawellek,2015PawKri}. Generally, the resolved extent of debris discs around lower luminosity stars are found to be greater than would be predicted from their blackbody temperature when compared to debris discs around higher luminosity stars. 

We calculate the parameter $\Gamma$ ($R_{\rm disc}/R_{\rm bb}$) for HD~105's disc using the relationship between stellar luminosity and disc temperature (radius) derived in \cite{2015PawKri}, tacitly assuming a dust composition of ice and astronomical silicate. We find $\Gamma = 2.43~\pm~0.42$, whereas a value of $4.94~\pm~0.42$ would have been predicted from the relationship shown in \cite{2015PawKri}. The dust grains that constitute the disc would thus be larger than we expect from \cite{2015PawKri}. 

However, the minimum grain size inferred from our revised SED model is 3.25$^{+3.00}_{-1.00}$~$\mu$m, consistent with the expectations of \cite{2015PawKri}. We also infer a size distribution exponent of $q = 3.55^{+0.30}_{-0.15}$. Our fit to the SED in the sub-millimetre is dominated by the far-infrared photometry, particularly the PACS 160~$\mu$m. In combination, the best-fit grain size and slopes parameters result in a steeper millimetre SED than inferred from the 9-mm ATCA data point. A relatively bright point source resides at 10\arcsec~from HD~105 in the ALMA image. Whilst this could be responsible for contaminating the APEX/LABOCA 880~$\mu$m data point, the ATCA beam was 5$\times$4\arcsec, so it would be well separated from the disc at 9~mm. Alternatively, stellar chromospheric emission could be contributing to the 9-mm flux measurement; enhanced emission (above that predicted from Rayleigh-Jean extrapolation) has been observed at millimetre wavelengths for several stars \citep[e.g.][]{2013Macgregor,2015Liseau,2016Chavez}. We might also infer that the dust grains have enhanced emissivity at sub-millimetre wavelengths, an inferrence supported by the relatively shallow millimetre spectral slope index $\beta = 0.3$. Increasing the emissivity of dust grains at millimetre wavelengths can be achieved in many ways, including increasing the porosity or adopting a core-mantle model.

As an additional data point in the menagerie of resolved debris discs around Sun-like stars, we can examine the properties of the system in contrast to discs around stars of similar spectral type. The Jena catalogue of resolved debris discs\footnote{\href{http://www.astro.uni-jena.de/index.php/theory/catalog-of-resolved-debris-disks.html}{http://www.astro.uni-jena.de/index.php/theory/catalog-of-resolved-debris-disks.html}} has 16 debris disc host stars that fit this criterion, taken from various literature sources \citep{2010Krist,2011Ertel,2012Krist,2013Eiroa,2014bMarshall,2014Soummer,2015Kennedy,2016DodsonRobinson,2016Choquet,2016LiemanSifry,2016Moor,2016Morales}. Amongst this cohort, the properties of HD~105's disc are fairly typical, having a radius of $\sim~90~$au (70 to 150 au), a typical dust temperature around 50~K (44 to 112 K), and a fractional luminosity of a few $\times10^{-4}$ (1$\times10^{-5}$ to 5$\times10^{-3}$). However, the constituents of this data set are heterogeneous, the discs having been resolved through various methods (optical, far-infrared, and millimetre) and the host stars having a wide variety of ages. 

Through angular averaging we have managed to obtain a detection of the disc in scattered light. The angular averaging method, used here for the first time, facilitated a detection of disc scattered light and a deeper probe of the dust albedo than obtainable from standard analysis of the imaging observations through injection of a disc model; application of the angular averaging method to similar cases of continuum bright discs that remain undetected in scattered light are expected to yield further detections. The architecture of HD~105's disc bears striking resemblance to those of both HD~207129 \citep{2010Krist,2011Marshall,2012Loehne} and HD~377 \citep{2016Steele,2016Choquet}. The albedo determined from the scattered light brightness calculated here is comparable to that found for dust grains in imaged discs of comparable fractional luminosity around nominally similar stars e.g. HD~207129 \citep[albedo $\sim$ 0.05][]{2010Krist}, and the properties of planetesimals in the Solar system.

\section{Conclusions}
\label{sec:con}

We have presented an analysis combining the interpretation of high resolution stellar spectra alongside new and archival scattered light, far-infrared, and (sub-)millimetre imaging observations of HD~105 and its circumstellar disc. In combination, these observations have provided the basis for a more detailed interpretation of the system, including its architecture and evolutionary state.

From a Bayesian approach applied to stellar evolution models we confirm its youth, with an age of {50~$\pm$~16~Myr}. Additionally, we found activity indicators which are consistent with being at a pre-main sequence evolutionary stage. The disc's planetesimal belt is spatially resolved at millimetre wavelengths. The best-fitting architecture is found to be a single annulus with a semi-major axis of 85~au and a width of 30~au at a moderate inclination of 50$\degr$. The disc is more compact than would be predicted from application of simple stellar luminosity-disc radius relationships, suggesting that the dust grains comprising the disc are either large, or efficient emitters at (sub-)millimetre wavelengths \citep[e.g. through increased porosity][]{2003delBurgo}.

There is no evidence of any offset between the planetesimal belt and the stellar position or any evidence of non-axisymmetric structure in the disc, either of which might imply the presence of a low mass companion interacting with the disc. No evidence of significant CO 2-1 line emission was found from the area of the disc, consistent with recent discoveries of gas in young ($<$ 60 Myr) debris discs being found in discs around A-type stars. 

Using an angular averaging technique to measure the scattered light radial profile, we have obtained a first detection of HD~105's debris disc in scattered light, determining a value for the dust albedo between 0.15 and 0.06. The application of this new technique to similar extant data sets, where the spatial extent of the disc is known from continuum emission but scattered light imaging has proven fruitless, should easily yield additional detections of debris discs in scattered light. This will provide an expanded pool of systems from which inferences can be drawn regarding their scattered light and thermal emission properties.

{Comparison of the dust albedo inferred from the scattered light with simple grain models and simplifying assumptions with water ice/astronomical silicate compositions allows us to infer that the dust grains are consistent with either pure astronomical silicate such as is typically assumed for debris dust, or a moderately icy composition as might be expected for material lying well beyond the snow-line of the system and consistent with SED modelling of a number of other debris discs.}

\acknowledgements

{We thank the anonymous referee for their constructive criticism that improved the manuscript.}

\textit{Herschel} is an ESA space observatory with science instruments provided by European-led Principal Investigator consortia and with important participation from NASA. 

This paper makes use of the following ALMA data: ADS/JAO.ALMA\#2012.1.00437.S. ALMA is a partnership of ESO (representing its member states), NSF (USA) and NINS (Japan), together with NRC (Canada) and NSC and ASIAA (Taiwan) and KASI (Republic of Korea), in cooperation with the Republic of Chile. The Joint ALMA Observatory is operated by ESO, AUI/NRAO and NAOJ.

This research has made use of the SIMBAD database, operated at CDS, Strasbourg, France. 

This research has made use of NASA's Astrophysics Data System.

{This paper has made use of the Python packages {\sc astropy} \citep{2013AstroPy,2018AstroPy}, {\sc SciPy} \citep{SciPy}, {\sc matplotlib} \citep{2007Hunter}, and {\sc Hyperion} \citep{2011Robitaille}.}

{JPM acknowledges research support by the Ministry of Science and Technology of Taiwan under grants MOST104-2628-M-001-004-MY3 and MOST107-2119-M-001-031-MY3, and Academia Sinica under grant AS-IA-106-M03.}

EC acknowledges support from NASA through Hubble Fellowship grant HF2-51355 awarded by STScI, which is operated by AURA, Inc. for NASA under contract NAS5-26555, for research carried out at the Jet Propulsion Laboratory, California Institute of Technology. This work is based on data reprocessed as part of the ALICE program, which was supported by NASA through grants HST-AR-12652 (PI: R. Soummer), HST-GO-11136 (PI: D. Golimowski), HST-GO-13855 (PI: E. Choquet), HST-GO-13331 (PI: L. Pueyo), and STScI Director's Discretionary Research funds. 

CdB acknowledges that this work has been supported by Mexican CONACyT research grant CB-2012-183007

GMK is supported by the Royal Society as a Royal Society University Research Fellow.

LM acknowledges support from the Smithsonian Institution as a Submillimeter Array (SMA) Fellow.

\bibliography{refs}

\end{document}